# Millisecond X–Ray Pulsar SAX J1808.4–3658: Limits to the Mass and the Radius of the Compact Star

Sudip Bhattacharyya[1,2]

[1]Joint Astronomy Program, Indian Institute of Science, Bangalore 560012, INDIA; sudip@physics.iisc.ernet.in
[2]Indian Institute of Astrophysics, Bangalore 560034, INDIA; sudip@iiap.ernet.in

## Abstract

We predict from a survey of equations of state and observations of X–ray pulsations from SAX J1808.4–3658, that the upper limit of the mass of the compact star is 2.27 $M_\odot$. The corresponding upper limit of the radius comes out to be 9.73 km. We also do a study to estimate the lower limit of the mass of the compact star. Such a limit is very useful to put constraints on equations of state. We also discuss the implications of the upper mass limit for the detection of the source in radio frequencies. We point out that the possible observation of radio–eclipse may be able to rule out several soft equation of state models, by setting a moderately high value for the lower limit of inclination angle.

## 1 Introduction

The discovery of millisecond X–ray pulsations (period $T = 2.49$ ms; Wijnands & van der Klis 1998) in the transient X–ray burster SAX J1808.4–3658 confirmed the speculation that LMXBs are progenitors of millisecond pulsars (Bhattacharya & van den Heuvel 1991). The orbital period ($P_{\rm orb} = 2.01$ hr) and the pulsar mass function ($f_1 = 3.7789 \times 10^{-5}$) of this source were observationally determined by Chakrabarty & Morgan (1998). These give valuable information about the masses (of both the primary and the secondary) and the inclination angle. For example, the value of $P_{\rm orb}$ uniquely determines the mass of a Roche lobe filling low mass star with known mass–radius relation.

It has been recently proposed that the compact star in SAX J1808.4–3658 is a strange star (SS) and not a neutron star (NS) (Li et al. 1999). Such a speculation, if confirmed, will prove that the so called *strange matter hypothesis* (Witten 1984) is correct. According to this hypothesis, strange quark matter (made entirely of deconfined $u$, $d$ and $s$ quark) could be the true ground state of strongly interacting matter rather than $^{56}$Fe. This is an important problem of the fundamental physics. To resolve it, we need to constrain the equations of state (EOS) for this compact star very effectively.

## 2 Limits to Mass and Radius

We estimate the upper limits of the mass and the radius of the compact star in SAX J1808.4–3658 by using the basic requirements for X–ray pulsations (if there is no 'intrinsic' pulse mechanism) and a criterion for the presence of accretion flow (that is not centrifugally inhibited). These give, adopting maximum to minimum observed flux ratio as 100 (see Li et al. 1999; Bhattacharyya 2001 for details),

$$R_1 < 7.40 m_1^{1/3} \text{ km}. \quad (1)$$

Here $m_1$ and $R_1$ are the mass (in unit of solar mass) and radius of the compact star respectively. Eqn. 1 gives the maximum value of $R_1$, if the maximum value of $m_1$ is known. To calculate $m_{1,\max}$ we first rewrite the eqn. 1 in the following form.

$$m_1 < 11.19 x_1^{-3/2}, \quad (2)$$

where $x_1$ is the dimensionless radius to mass ratio of the compact star. We can compute $m_{1,\max}$ from eqn. 2, if the minimum value of $x_1$ is known. To choose the value of $x_{1,\min}$, we survey about 20 EOS (that include both SS and NS) and examine the value of $x_1$ corresponding to the maximum possible mass for a given EOS. For both SS and NS, we choose EOS of widely varying stiffness parameters, which guarantees our results to be of sufficient generality. This is reflected by the wide range of maximum possible mass values given in Table 1, where we have listed 13 representative EOS. From Table 1 (and Figure 1 of Bhattacharyya 2001), we notice that the $x_1$ values for all the EOSs are confined to the range $2.98 - 4.34$, with 11 (out of 13) points clustering in $3.3 - 3.7$. As for none of the EOS models, $x_1$ value is less than 2.9, we take this value as the lower limit of $x_1$. Such a conclusion is very general, as it is valid for the whole range of existing EOS models. This gives 2.27 as the upper limit of $m_1$ from eqn. 2. The corresponding upper limit of $R_1$ comes out to be 9.73 km from eqn. 1.

If we write the well-known expression for the pulsar mass function ($f_1$) as

$$\sin i = f_1^{1/3} \frac{(m_1 + m_2)^{2/3}}{m_2} \quad (3)$$

we see that for a given value of $m_{2,\min}$ (minimum value of companion's mass in unit of solar mass), if $i_{\min}$ (minimum value of inclination angle) is greater than a certain value, every $i_{\min}$ will correspond to a minimum possible value of $m_1$. Therefore if we can observationally constrain $i$ from the lower side, the value of $m_{1,\min}$ can be predicted. For example, the detailed modeling of the optical companion's multiband photometry during outburst with a simple X–ray heated disk model suggests that $\cos i < 0.45$ (Wang et al. 2001, in preparation; Bildsten & Chakrabarty 2001) for SAX J1808.4–3658. This implies $i > 63^o$ and hence $m_{1,\min} = 1.48$

## 3  Discussions

In this paper, we have estimated the upper limits of the mass and the radius of the compact star in SAX J1808.4-3658. Li et al. (1999) have concluded that a narrow region in $m_1 - R_1$ space will be allowed for this star. The upper boundary of the mass will constrain this region effectively. It can also give the upper limit of $i$ (from eqn. 3), if $m_{2,\min}$ is known by an independent measurement. Alternatively, $m_{1,\max}$ gives the upper limit of $m_2$, for a known value of $i_{\min}$. For example, $i_{\min} = 63^o$ gives $m_{2,\max} = 0.066$ for $m_{1,\max} = 2.27$.

It has been proposed that SAX J1808.4-3658 may emerge as a radio pulsar during the X–ray quiescence (Chakrabarty & Morgan 1998). Ergma & Antipova (1999) have calculated that for $\lambda < 3$ cm, it may be possible to observe radio emission from this source. However our limits of mass values give a slightly higher (3.8 cm) upper limit for $\lambda$.

We have already mentioned in the previous section that a moderately high value of $i_{\min}$ will give a lower limit of $m_1$. This will be very important for constraining EOS more decisively (as corresponding to every EOS, there exists a maximum possible mass). For example, if $i_{\min} = 63^o$ (corresponds to $m_{1,\min} = 1.48$, given in the previous section), our EOS models SS1, SS2 and Y will be unfavoured (see the '$m_{1,\max}$' column of Table 1). According to Chakrabarty & Morgan (1998), a deeper eclipse might be observed for the less penetrating radio emission, providing a strong constraint on the value of $i$. Therefore, we expect that, the value of $i_{\min}$ (determined by this method) may be able to rule out several soft EOS models in future.

Table 1: List of 13 EOS (both SS and NS) of widely varying stiffness parameters and their references. The values of relevant properties (see the text) are also given.

| EOS label | compact star | Reference | $m_{1,\max}$ | $R_1(km)$ | $x_1$ |
|---|---|---|---|---|---|
| SS2 | SS | Dey et al. (1998) | 1.32 | 6.53 | 3.34 |
| SS1 | SS | Dey et al. (1998) | 1.44 | 7.07 | 3.32 |
| $B_{90}$ | SS | Farhi & Jaffe (1984), $B = 90$ MeV/fm$^3$, $m_s = 0$ | 1.60 | 8.74 | 3.69 |
| $B_{60}$ | SS | Farhi & Jaffe (1984), $B = 60$ MeV/fm$^3$, $m_s = 0$ | 1.96 | 10.71 | 3.70 |
| Y | NS | Pandharipande (1971b), hyperonic matter | 1.41 | 7.10 | 3.39 |
| B | NS | Baldo, Bombaci, & Burgio (1997), nuclear matter | 1.79 | 9.64 | 3.64 |
| W | NS | Walecka (1974), neutron matter | 2.28 | 11.22 | 3.32 |
| SBD | NS | Sahu, Basu, & Datta (1993), nuclear matter | 2.59 | 14.08 | 3.68 |
| A | NS | Pandharipande (1971a), Reid soft core | 1.66 | 8.37 | 3.42 |
| AU | NS | Wiringa, Fiks, & Fabrocini (1988), AV14 + UVII | 2.13 | 9.41 | 2.98 |
| FPS | NS | Lorenz, Ravenhall, & Pethick (1993), UV14 + TNI | 1.80 | 9.28 | 3.48 |
| L | NS | Pandharipande & Smith (1975b), mean field | 2.70 | 13.70 | 3.43 |
| M | NS | Pandharipande & Smith (1975a), tensor interaction | 1.81 | 11.60 | 4.34 |